# Topological surface state in the Kondo Insulator Samarium Hexaboride

D. J. Kim, J. Xia, and Z. Fisk

Department of Physics and Astronomy, University of California, Irvine, Irvine, California 92697, USA

**Strongly correlated electron systems show many exotic properties such as unconventional superconductity, quantum criticality, and Kondo insulating behavior. In addition, the Kondo insulator $SmB_6$ has been predicted theoretically to be a 3D topological insulator with a metallic surface state. We report here transport measurements on doped $SmB_6$, which show that ~3% magnetic and non-magnetic dopants in $SmB_6$ exhibit clearly contrasting behavior, evidence that the metallic surface state is only destroyed when time reversal symmetry is broken. We find as well a quantum percolation limit of impurity concentration which transforms the topological insulator into a conventional band insulator by forming impurity band. Our careful thickness dependence results show that $SmB_6$ is the first demonstrated perfect 3D topological insulator with virtually zero residual bulk conductivity.**

Topological invariants of electron wave functions in condensed matter physics reveal many intriguing phenomena (1,2). The most exotic one is the topological insulator (TI) characterized by the $Z_2$ group where an insulating bulk coexists with a metallic boundary state (3,4). Possible novel quantum states supporting coherent qubits using Majorana fermions with their potential for technological application has led to intense research into Bi based TIs (5). Their large band gap and simple surface state make it possible to explore their underlying physics using various techniques (6). However, the main complication concerning these Bi based materials is that they still have considerable residual conductivity in the bulk, and only experimental techniques distinguishing bulk and surface clearly such as angle-resolved photoemission spectroscopy (ARPES) (7,8) or scanning tunneling microscope (STM) (9,10) can be used to explore the surface properties of these materials properly. Thus, it is both not easy to detect their topological surface state with transport techniques and hard to achieve scalability in applications based on transport properties. Theories (11,12) predict that a Kondo insulator $SmB_6$, which evolves from a Kondo lattice metal to an insulator with a small gap as the temperature is lowered, could be a topological insulator with a metallic surface state. The insulating bulk and metallic surface separation has been demonstrated in recent transport measurements (13-15). This bulk and surface separation is especially important for potential applications toward scalable quantum information processing (16).

The Kondo insulator $SmB_6$ is a dense lattice of Sm magnetic moments which evolves from a dirty metal at room temperature to an insulator with some residual conductance at low temperature (17). This transition is one of the most remarkable phenomena of Kondo lattices which also exhibit unconventional superconductivity (18), hidden order transition (19) and quantum criticality (20) as a result of immersion of magnetic moments in a conduction band. The resistance of $SmB_6$ increases exponentially as the temperature decreases (see Fig. 1A) with a non-universal ratio of low to high temperature resistance. Usually, the higher quality sample exhibits the higher ratio (see Supplementary Materials).

From Ohm's law, the electrical resistivity of a rectangular parallelepiped shaped bulk conductor is defined by the product of measured resistance and geometrical factor L / A, where L and A are the length and cross sectional area. Since ideal TIs do not have bulk conductance, they cannot satisfy the basic Ohm's law. Thus, if a 3D TI transforms from a conventional bulk conductor at high temperature to an insulator with only surface conduction at low temperature, the sample thickness should not affect the measured low temperature limiting resistance but be independent of it. Figure 1B shows this most unusual behavior of $SmB_6$ manifesting the thickness independence of sample resistance caused by the bulk and surface separation is a necessary condition for an ideal TI. The sample thickness is reduced by polishing. After the first resistance versus temperature measurement, the same sample is flipped and mounted on a polishing fixture to reduce the thickness from the backside to keep the original surface and electrical leads made by spot welding intact. (right lower inset Fig. 1A and see Supplementary Materials). The resistance is measured in a Quantum Design PPMS with LR700 ac-bridge and automation software (21). From room temperature down to 10 K, the comparative resistance ratio (RR) of three different thicknesses of $SmB_6$ follows the geometric ratios as with usual bulk metallic systems, but below 10 K, RR starts dropping and rapidly converges to one below 5 K. This convergence indicates that the surface conductance dominates with a relatively much small conductance being left from the finite temperature effect coming from the tiny band gap in the bulk, which indicates complete insulating bulk and metallic surface separation at low temperature. These temperatures are consistent with our previous reports (13-14). In contrast to $SmB_6$, semiconducting $BaB_6$ and quantum critical $CeAuSb_2$ (22) show only the bulk metallic behavior over entire temperature range. However, this separation of bulk and surface is not enough to claim that $SmB_6$ is indeed a topological Kondo insulator (TKI).

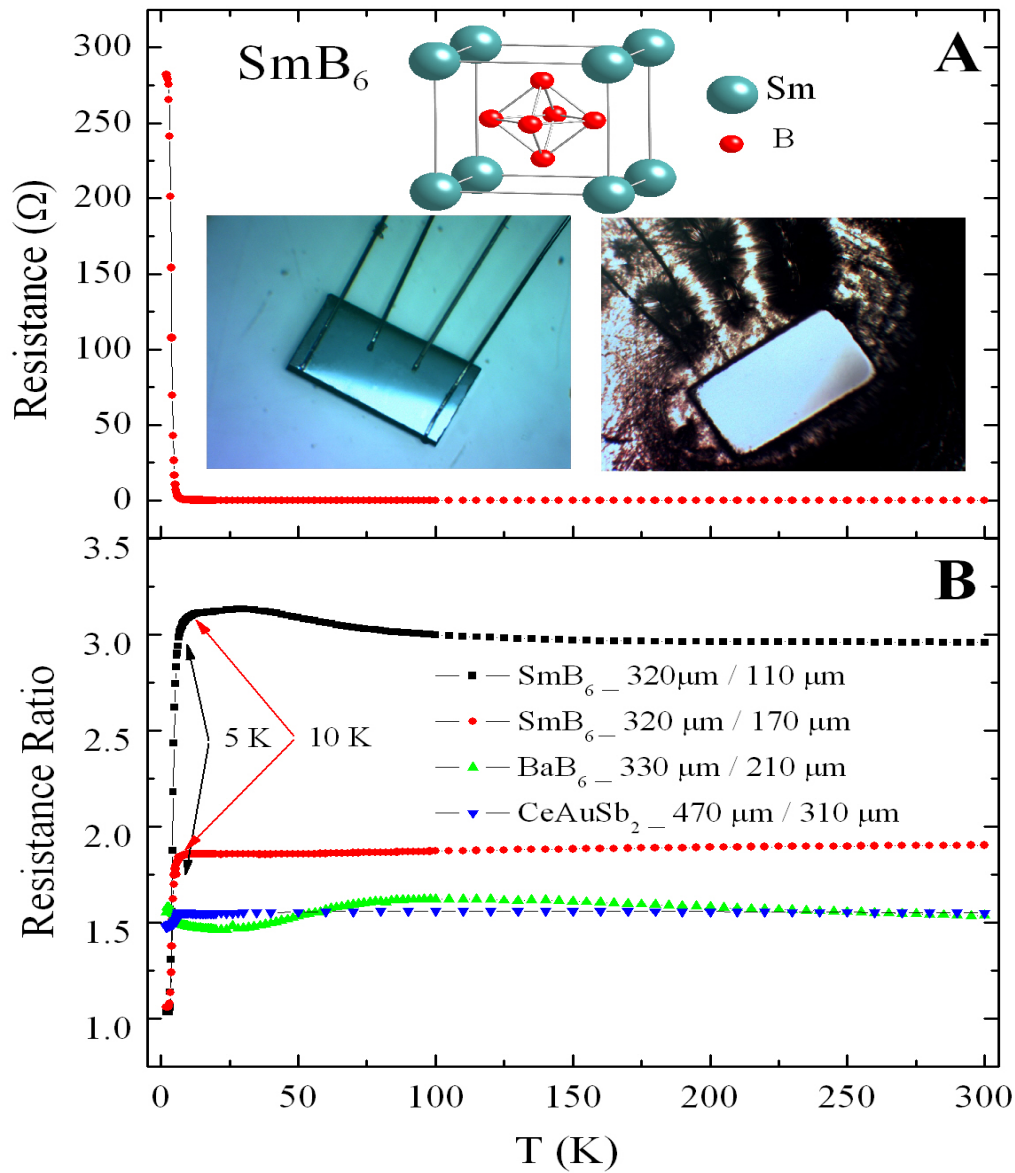

**Fig. 1. Failure of Ohm's law.** (**A**) Resistance versus temperature of rectangular parallelopiped shaped $SmB_6$ sample, the insets show crystal structure (upper), a finely polished surface sample with electrical leads (lower left), and the sample flipped and mounted on a polishing fixture for thickness reduction with the original leads in place (lower right). (**B**) thickness dependences of $SmB_6$, $BaB_6$ and $CeAuSb_2$ resistance. $SmB_6$ has very clear thickness independence and its resistance ratios for three different thicknesses converge to one indicating bulk (insulator) and surface (conductor) separation as the temperature is lowered below 10 K. In contrast, $BaB_6$ and $CeAuSb_2$ show conventional bulk conducting behavior.

The most convincing way to prove $SmB_6$ to be a TKI is with ARPES to observe the Dirac cones. However, if SmB6 is an ideal TKI with virtually zero bulk conductance at low temperature, it should be possible to probe with transport measurement not only for understanding underlying physics but also for hybrid system with other superconducting and magnetic materials. Even though the thickness independence is exotic, it could arise from an accidental layer on the surface (23). Thus, it is necessary to find firmer evidence to support surface Dirac fermions along with the thickness independent resistance at low temperature. Usually, to probe the surface Dirac fermion, quantum oscillations from Berry phase and Landau quantization by STM or magneto-optics experiments have been used for the conventional TIs (4-6). However, the Kondo insulator is a strongly correlated system with the small band gap arising from Kondo screening and magnetic field may affect not only the nontrivial band topology but also quench the Kondo effect. The previous magneto resistance measurement up to 9 T did not show any evidence of quantum oscillation (13) and even higher field is reported to quench the Kondo insulating property leading to more metallic behavior (24). Therefore, conventional transport measurements in high magnetic field may not be the way to detect Dirac fermions through quantum oscillation. Instead, TIs have three aspects of the topological protection of the surface state (4-6,9,10). First, its fundamental $Z_2$ topology preserves a gapless surface state unless time reversal symmetry (TRS) is broken. Second, helical spin polarization prevents momentum backscattering from k to –k by non-magnetic impurities. Finally, the Berry phase protects the surface state from weak localization through time reversed paths. These collectively provide a robust surface state with TRS conservation. Probably, the simplest manifestation of this TRS protected surface state would be positive magneto-resistance in an external magnetic field, and a transition from positive to negative would be expected for TKI from the above reasoning. Indeed this happens for pure $SmB_6$ below 1 K, but the observation of this transition does not uniquely support a topological surface state (see Supplementary Materials).

To include the effect of the broken TRS, we use magnetic and non-magnetic impurity doping in $SmB_6$ and measure resistance versus temperature with the thickness reduction method to check if the low temperature resistance value converges to one as the pure samples do. Figure 2A shows a clear contrast between magnetic gadolinium (Gd) and non-magnetic Yttrium (Y) and Ytterbium (Yb) doped on Sm sites in $SmB_6$ samples. The overall RRs for two different thicknesses in Yb and Y doped $SmB_6$ exhibit bulk conducting property at high temperature which converge to one at low temperature. In contrast, the RR of the Gd doped sample follows the geometrical ratio over the entire temperature range. Usually both magnetic and non-magnetic doping can break the Kondo insulating phase and make doped samples quite metallic at concentration higher than 30 %, eventually leading to antiferromagnetic ($GdB_6$), conventional semiconducting ($YbB_6$) and superconducting transition ($YB_6$). In the intermediate doping concentration, the resistance rise at low temperature tends to be decrease dramatically. Thus, the contrasting RRs or bulk conducting behavior for the Gd doped sample could be misinterpreted rather as a more effective reduction of the resistance rise at low temperature. However, as shown in Fig. 2B and 2C, compared with the Gd doped sample which makes almost 2.5 orders of magnitude jump, the Y doped sample makes less than 2 orders of magnitude increase. Remarkably, below 4 K, the resistance of the Y doped sample saturates but the Gd doped sample shows further dramatic increase even to the base temperature. We can confirm from other samples in the same batches that the Gd doping makes a dramatic resistance rise compared with Y doping (see Fig. 2B and 2B insets) in the temperature regime from 20 K to 500 mK, where the resistance increases ( R(T)/R(20 K) ) are 237 and 23 for Gd doping and Y doping respectively.

Substituting Sm with Y or Gd decreases the resistance significantly but still there survives a band gap from the slightly disturbed hybridization at ~3% doping concentration. Thus, the Y doped sample's RR convergence and low temperature saturation is consistent with the prediction for much more conductive surface state, but when TRS is broken by the magnetic impurities, the bulk resembles very tiny gap insulator with some thermally induced conduction at finite temperature while the resistance keeps increasing with lowering temperature due to the absence of metallic surface state. This result is a required property of a topological surface state protected by TRS invariance.

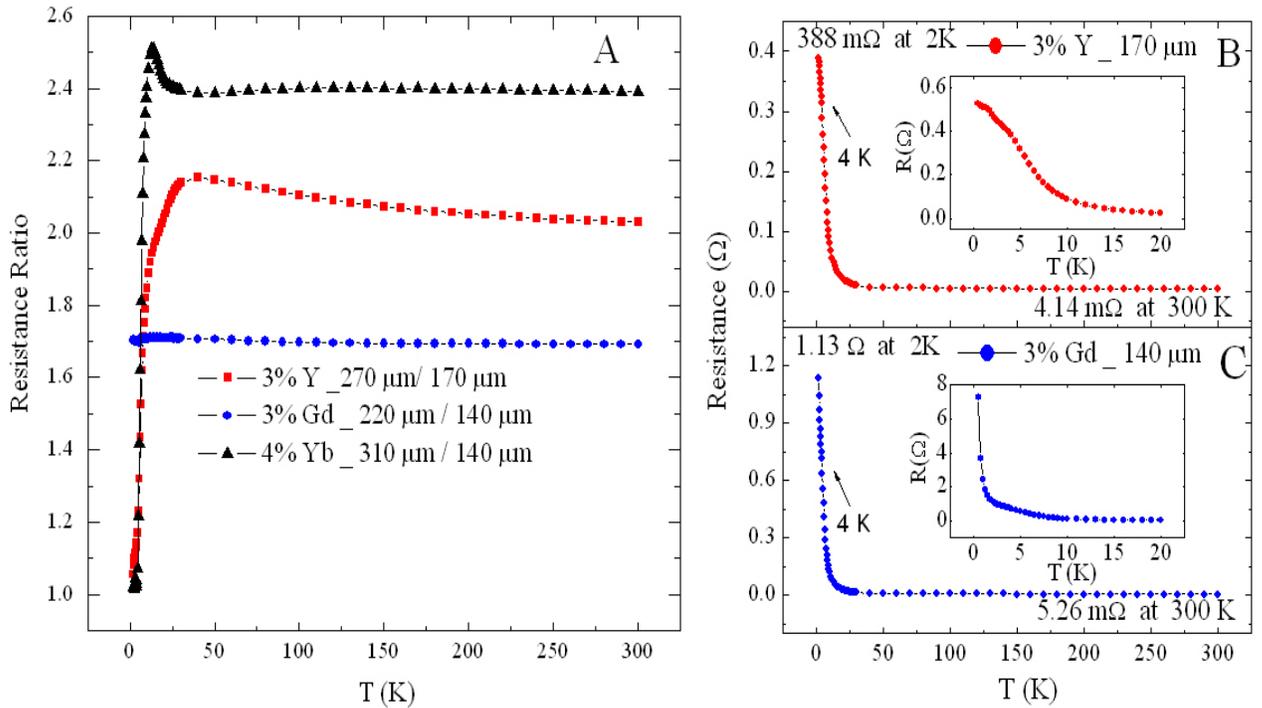

**Fig. 2. Topologically surface state protected by time reversal symmetry. (A)** Thickness resistance ratios for Y (3%), Gd (3%), and Yb (4%) doped $SmB_6$ samples. In contrast to the non-magnetic impurities which do not destroy the surface state, the magnetic impurities make the sample a conventional insulator. **(B-C)** Resistance versus temperature curve of Y doped and Gd doped $SmB_6$. The Gd doped sample makes larger resistance rise and does not show resistance saturation at low temperature. Insets in **(B)** and **(C)** show resistance down to 500 mK for each doping.

An important question concerning impurities on Sm sites is whether or not they are dynamically coupled to the conduction electrons and involved in the Kondo insulator formation in the bulk. If they participate in a more complex way or generate another competing order, the interpretation of Fig. 2 might be more subtle. The surface state arises from the topological property of the bulk. Thus, the ideal doping condition is to keep the Kondo insulating property intact and introduce just enough impurities to demonstrate the effect on the surface state. Figure 3A shows the inverse of magnetic susceptibility obtained by subtracting $\chi$ of 3% Y doped sample from that of 3% Gd doped sample to elucidate the pure Gd behavior at 1000 Oe. The inverse susceptibility for the Gd doped sample exhibits a straight line passing through the origin which indicates that the magnetic impurities obey a simple paramagnetic Curie law. But 40% Gd leads the sample beyond the percolation limit and the Curie law behavior disappears at low temperature (Fig. 3B) in contrast to the 3% Gd doping (upper inset, Fig. 3B). The magnetization curve (lower inset, Fig. 3B) suggests antiferromagnetic interactions between Gd moments. To confirm the non-interacting impurity picture in the low concentration doped samples, we measured the magnetization for both Gd and Y doped samples and subtracts again to get Brillouin function fits up to 2 T perfectly as shown in Fig. 3C-E. Higher field alters the Kondo effect in the $SmB_6$ host and changes the free Gd response from its ideal paramagnetic behavior.

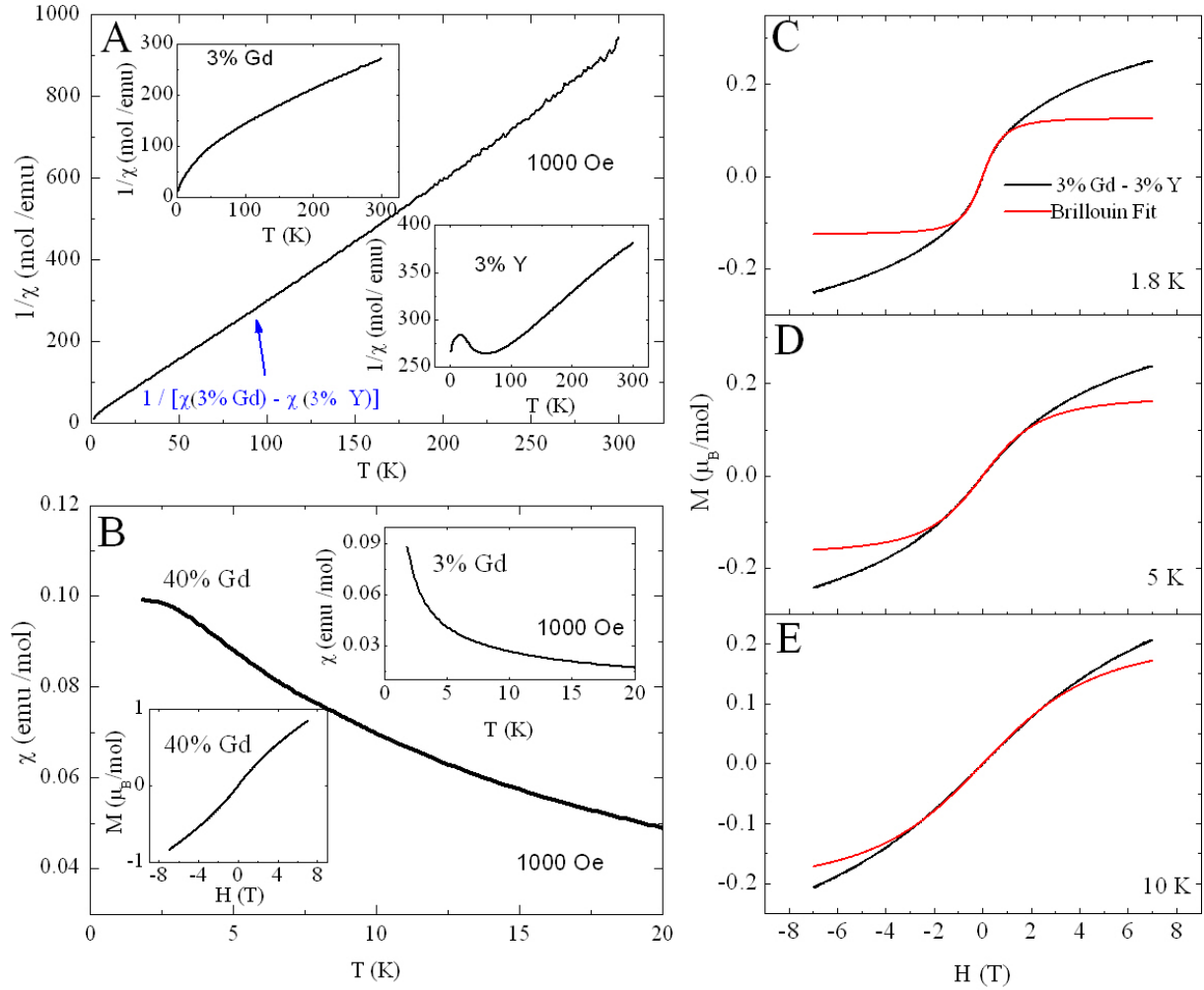

**Fig. 3. Behavior of magnetic impurities in SmB$_6$.** **(A)** Inverse magnetic susceptibility of subtracted Gd contribution with temperature obeying Curie's law. Insets, inverse susceptibility of 3% Gd (upper) and 3% Y (lower). **(B)** Magnetic susceptibility of 40% Gd doped sample shows saturation below 3 K, magnetization curve versus field for the sample has a straight line portion (lower inset) indicating possible magnetic interaction between Gd sites at lower temperature. In contrast, the susceptibility of low concentration (3%) sample exhibits divergence at low temperature (upper inset). **(C-E)** Magnetization versus field for subtracted 3% Gd portion at 1.8 K, 5 K, and 10 K. the red lines are Brillouin fits to each curves.

Independent of dopant type (magnetic or non-magnetic), a high concentration of impurities in a Kondo insulator can affect its properties. Substituting for Sm with impurities decreases the resistance significantly and large doping (above 30%) destroys the insulating properties and leads to a metallic state (25). When the doping concentration is modest and lies between the strong Kondo insulating state and bulk metallic state, the surface state can be altered even with non-magnetic impurities. Divalent Yb does not have magnetic moments in $SmB_6$ (see Supplementary Materials). The resistance of a 4% Yb doped sample has an almost 4 order of magnitude rise with lowering temperature and very clear bulk and surface separation with concentration (see, Fig. 4 A, B). However, the RR for an 18% doped sample does not have the bulk and surface separation characteristic (Fig. 4A). As with the 3% Gd and Y doped samples in Fig. 2, figure 4 B and C show clear difference at low temperature. The 4% Yb doped sample shows the usual Kondo insulating surface state with saturating resistance but 18% Yb instead has a diverging feature in the low temperature range. This increasing resistance indicates a bulk insulating property and this is probably connected with the fact that pure $YbB_6$ is a typical band semiconductor

A single impurity or defect behaves as a boundary in the system. As the $Z_2$ index varies at the boundary, localized in-gap bound states with opposite spins with the gapless Dirac dispersion can be formed in topological insulators and the wave function decays exponentially away from the center of the impurity with a characteristic length scale. When the length scale is smaller than the distance among impurities, the energy overlap of these bound states leads to possible quantum scatterings between them at each impurity site (26). But when the impurity concentration is low enough, this quantum scattering is prohibited to ensure the robustness of the surface state. Thus, as the non-magnetic impurity concentration increases, the scattering in the dense bound states can form an extra impurity band of the overlapping bound states, and this makes it possible to induce back scattering to destroy the surface state. Thus, the possible insulating state in fig. 4C can be in the quantum percolation limit where a transition from a topological insulator to a conventional insulator occurs.

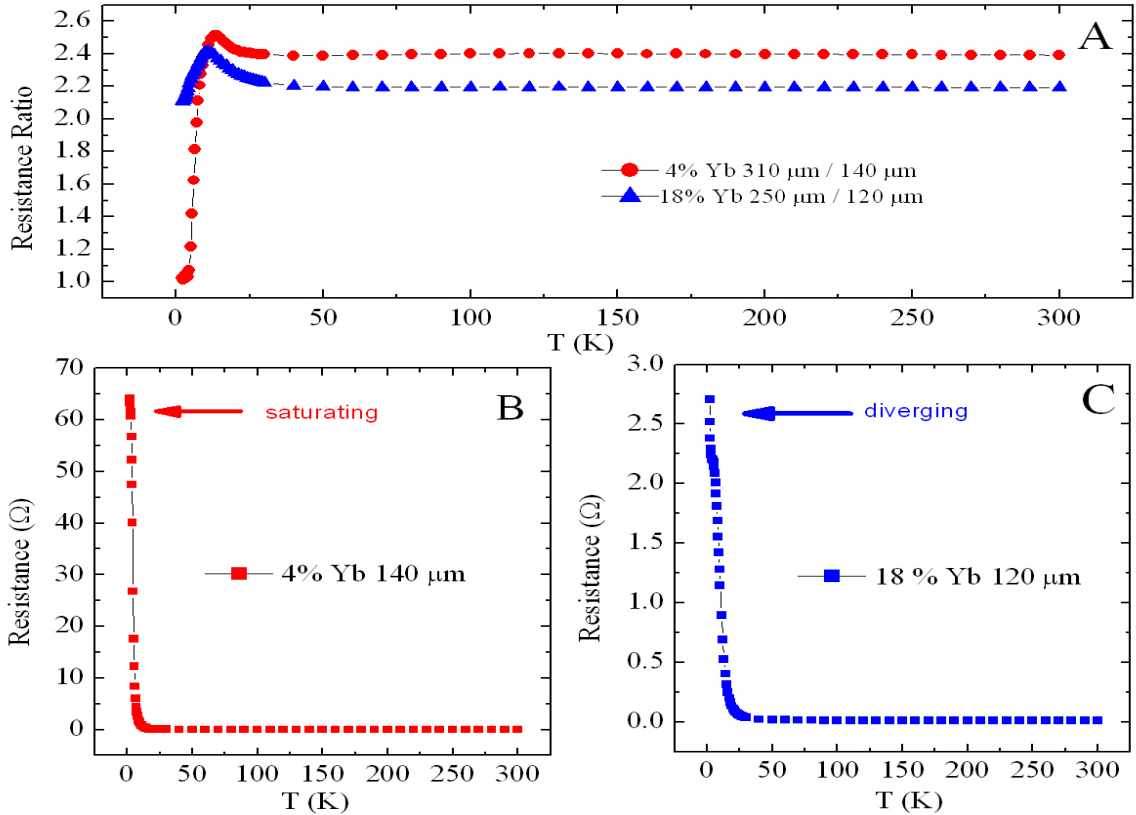

**Fig. 4. Quantum percolation of high concentration Yb doped SmB$_6$.** **(A)** Thickness resistance ratios for 4% Yb and 18% Yb doped SmB6 samples. **(B-C)** Resistance versus temperature curves for each concentration. At low temperature 18% doping resistance diverges when compared with 4% doping. This indicates a transition to a conventional insulator by forming impurity band above the percolation limit.

**Acknowledgements**

We thank M. Dzero, H. Lee, H. K. Lee, S. Thomas, I. Krivorotov, and W. Ho for discussions. This research was supported by NSF-DMR-0801253 and UC Irvine CORCL

Grant MIIG-2011-12-8. D.J.K and Z.F conceived the experiment idea and grew crystals. D.J.K performed the measurements. All authors discuss the results and participate in data analysis. D.J.K and Z.F wrote the manuscript.

**Supplementary Materials**

[www.sciencemag.org](www.sciencemag.org).

Supplementary Text

Figs. S1 to S3

# Supplementary Materials for

# Topological surface state in the Kondo Insulator Samarium Hexaboride

D. J. Kim, J. Xia, and Z. Fisk

Department of Physics and Astronomy, University of California, Irvine, Irvine, California 92697, USA

## S1. Sample quality and thickness reduction

We used Al flux growth in continuous Ar purged vertical high temperature tube furnace with high purity elements to grow all single crystals. The samples are leached out in sodium hydroxide solution.

The key feature of $SmB_6$ resistance is an exponential rise with cooling with low temperature saturation. However, the order of magnitude, saturation rate, and saturation point of resistance rise are not sample independent for samples from different batches. Usually higher purity elements (Sm, B) including flux material, Al, make larger increase with slightly lower saturation temperature (see Fig. S1 A).

The high quality sample and thickness reduction with well defined rectangular parallelipiped geometry are two important requirements for the resistance ratio measurement. We choose samples with well defined facets for reference and shape them with $Al_2O_3$ polishing pads (usually start from 30 μm and end with 50 nm roughness pad) into parallelipipeds. Then, we used dilute HCl (50 HCL + 50 DI water) for 2 minutes to remove possible oxidation on the surface. After drying the sample, the electrical leads are made with thermocouple grade 25 μm thick platinum wires by spot welding. After measuring the thickness with microscope reticule, the resistance temperature dependence is precisely measured at each stabilized temperature with a high resolution AC resistance bridge and Quantum design PPMS. One important thing is that the thermal cycle (cooling down and warming up), mounting sample on the polishing fixture with wax, and following chemical treatment (Acetone, IPA, HCL etching) should not change the resistance dependence with temperature for the same thickness sample to avoid any possible artifact from measurement setup (drift, offset, thermal cycling effect) and chemical reaction or contamination (see Fig. S1B). Measuring resistance versus temperature for one thickness, the reduced thickness data is measured with the exactly same sample with the original leads. First, the same sample with the original leads is flipped to mount on the fixture, and the leads are carefully pressed so as not to be destroyed during the polishing. Figure S1C shows the procedure for thickness reduction.

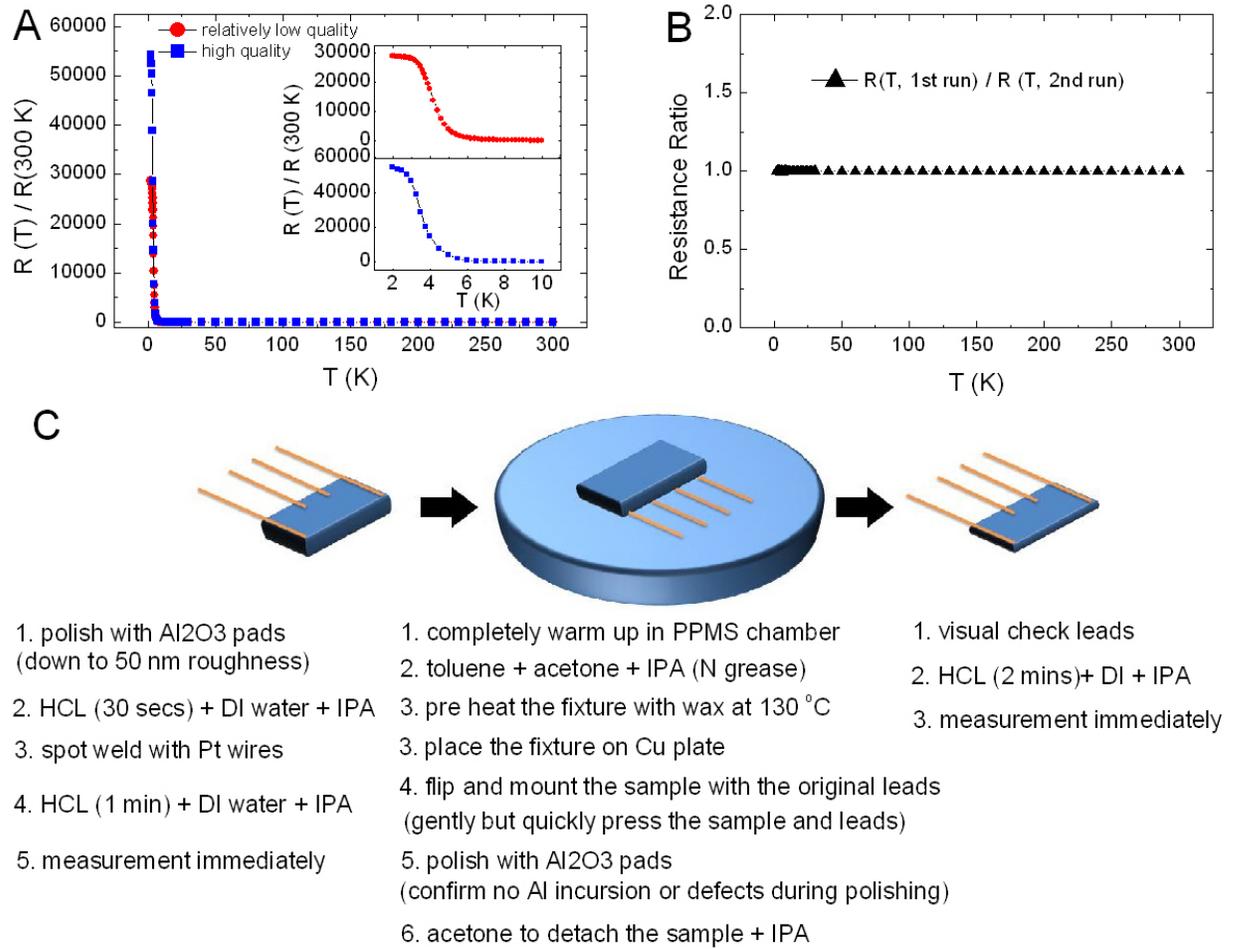

**Fig. S1. Sample quality, reproducibility, and thickness reduction.** (**A**) resistance versus temperature curves from two different batches. (**B**) resistance ratio versus temperature, the first run is measured with the initial condition of **c** and the second run is measured with the final condition of **C** without polishing. (**C**) sample thickness reduction procedure.

**S2. Transition from positive to negative magneto-resistance**

Usually f orbital Kondo insulators show negative magneto resistance associated with Kondo screening breaking and eventually evolve to positive when they become a more metallic bulk conductor at higher temperature. One possible conjecture for TKI would be a sign change of magneto resistance with field, from positive at low field to negative at high field. This happens in high purity SmB6 samples (see Fig. S2A). However, this transition is not unique to temperatures below complete gap opening temperature. Even above 10 K with magnetic or non-magnetic impurity doped the samples show a similar transition (see Fig. S2B). Thus this field dependence is not clear support for the TKI state.

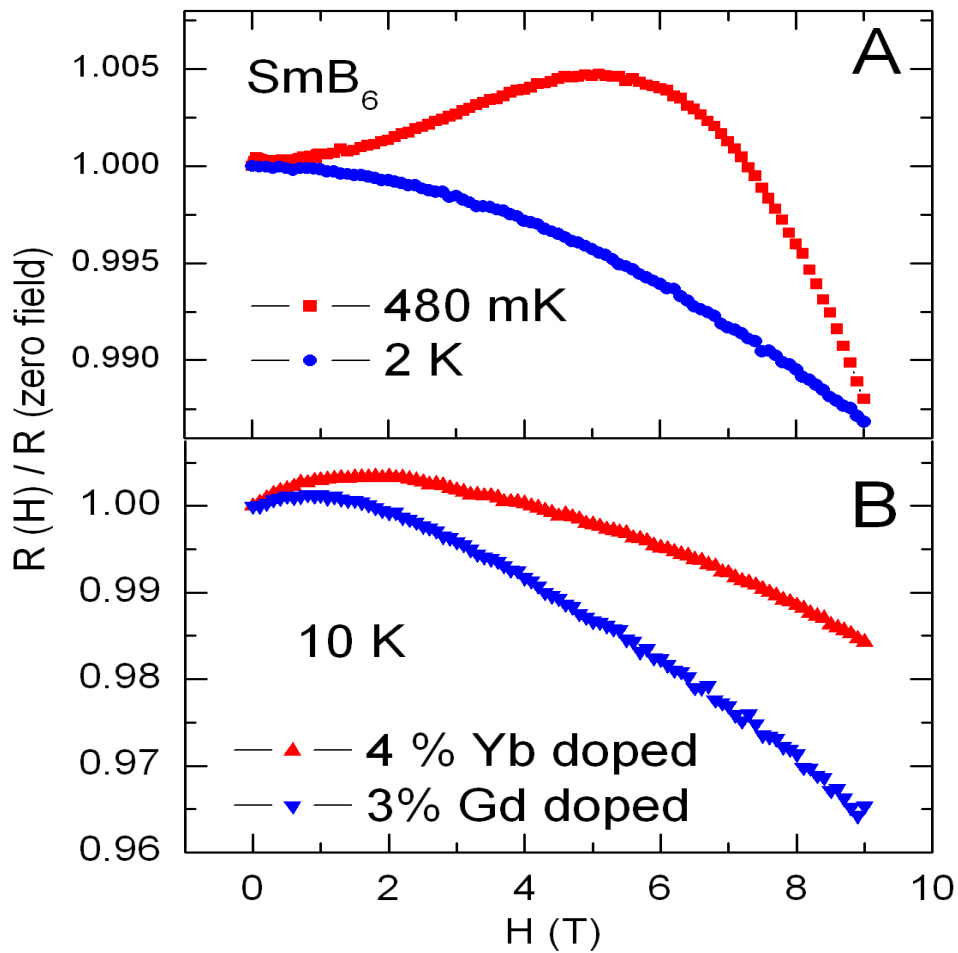

**Fig. S2. Sign change of magneto-resistance.** (**A**) resistance versus external magnetic field curves for pure $SmB_6$ at 480 mK and 2 K. (**B**) resistance versus external field of doping samples at 10 K.

## S3. Magnetic property of Yb doped $SmB_6$

Impurities with magnetic moments on Sm sites usually contribute to the magnetic properties of the doped samples significantly in magnetic susceptibility and magnetization curve, as in Gd doped samples. We find that divalent Yb, which has magnetic moment in free ion form, does not show a magnetic moment in $SmB_6$ and does not disturb the insulating properties in host bulk as do Y and Gd impurities. Figure S3 shows magnetic susceptibility curve for 18% Yb doped $SmB_6$ at 1000 Oe and magnetization curve (inset) compared with pure $SmB_6$. Yb sites appear like a magnetic vacancy in those curves,.The non magnetic doping shifts the peak and valley positions as in other non-magnetic doped samples.

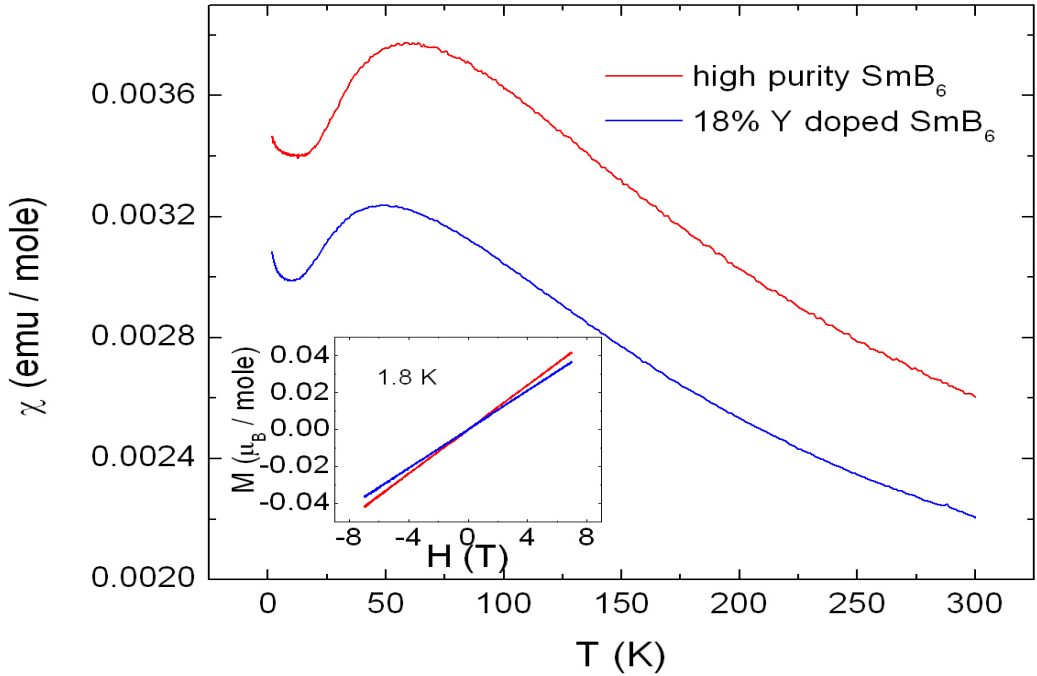

**Fig. S3. Magnetic properties of Yb doped $SmB_6$.** Magnetic susceptibility (1000 Oe external field) and magnetization (inset) curves of high purity pure and 18% Yb doped $SmB_6$.